\documentclass[11pt,a4paper]{article}

\usepackage[utf8]{inputenc}
\usepackage[T1]{fontenc}
\usepackage{newunicodechar}
\usepackage[margin=1in]{geometry}
\usepackage{amsmath}
\usepackage{amssymb}
\usepackage{stmaryrd}
\usepackage{booktabs}
\usepackage{array}
\usepackage{tabularx}
\usepackage{authblk}
\usepackage{xcolor}
\usepackage{graphicx}
\usepackage{longtable}
\usepackage{calc}
\usepackage{listings}
\usepackage[colorlinks=true,linkcolor=blue!40!black,citecolor=blue!40!black,urlcolor=blue!40!black]{hyperref}
\usepackage{xurl}
\lstdefinestyle{leanbox}{
  basicstyle=\small\ttfamily,
  columns=flexible,
  breaklines=true,
  keepspaces=true,
  showstringspaces=false,
  extendedchars=false,
  inputencoding=utf8,
}
\newunicodechar{∈}{\ensuremath{\in}}
\newunicodechar{≤}{\ensuremath{\leq}}
\newunicodechar{≥}{\ensuremath{\geq}}
\newunicodechar{≠}{\ensuremath{\neq}}
\newunicodechar{μ}{\ensuremath{\mu}}
\newunicodechar{Σ}{\ensuremath{\Sigma}}
\newunicodechar{⟨}{\ensuremath{\langle}}
\newunicodechar{⟩}{\ensuremath{\rangle}}
\newunicodechar{∪}{\ensuremath{\cup}}
\newunicodechar{∩}{\ensuremath{\cap}}
\newunicodechar{∖}{\ensuremath{\setminus}}
\newunicodechar{∅}{\ensuremath{\emptyset}}
\newunicodechar{→}{\ensuremath{\rightarrow}}
\newunicodechar{←}{\ensuremath{\leftarrow}}
\newunicodechar{∀}{\ensuremath{\forall}}
\newunicodechar{∃}{\ensuremath{\exists}}
\newunicodechar{∧}{\ensuremath{\wedge}}
\newunicodechar{∨}{\ensuremath{\vee}}
\newunicodechar{↔}{\ensuremath{\leftrightarrow}}
\newunicodechar{∉}{\ensuremath{\notin}}
\newunicodechar{≡}{\ensuremath{\equiv}}
\newunicodechar{∑}{\ensuremath{\sum}}
\newunicodechar{∞}{\ensuremath{\infty}}
\newunicodechar{⌊}{\ensuremath{\lfloor}}
\newunicodechar{⌋}{\ensuremath{\rfloor}}
\newunicodechar{∎}{\ensuremath{\square}}
\newunicodechar{⇒}{\ensuremath{\Rightarrow}}
\newunicodechar{φ}{\ensuremath{\varphi}}
\newunicodechar{ε}{\ensuremath{\varepsilon}}
\newunicodechar{π}{\ensuremath{\pi}}
\newunicodechar{Ω}{\ensuremath{\Omega}}
\newunicodechar{⁻}{\ensuremath{{}^{-}}}
\newunicodechar{₀}{\ensuremath{_0}}
\newunicodechar{₁}{\ensuremath{_1}}
\newunicodechar{₂}{\ensuremath{_2}}

\title{\textbf{Revisiting the Average Case Complexity of Multilevel Syllogistic}}

        \author[1]{\textbf{Lars Warren Ericson}}
        \affil[1]{Catskills Research Company}
        \affil[1]{\texttt{lars.ericson@catskillsresearch.com}}

        \date{\today}

\begin{document}

        \maketitle

        \begin{center}
          \small
          \textbf{ORCID:} 0000-0001-8299-9361 \\
          \textbf{Primary Category:} cs.LO (Logic in Computer Science) \\
          \textbf{Secondary Category:} cs.CC (Computational Complexity) \\[0.5em]
          \textbf{Lean~4 formalization:} \url{https://github.com/catskillsresearch/avg\_case\_mls}
        \end{center}

        \begin{abstract}
        We describe a Lean~4 formalization revisiting NYU Courant Technical Report TR1995-711
on the average-case complexity of Multilevel Syllogistic (MLS). The development encodes
Reischuk--Schindelhauer average-case classes, an axiomatic MLS/EMLS semantics layer, a
partial Ferro--Omodeo--Schwartz decision procedure with proved soundness and partial
completeness on a membership-free fragment, serialization and step budgets, and
conditional NP-average completeness and non-AvP hardness corollaries modulo explicitly
documented structural axioms. Full Lean sources are inlined in the appendix modules.
        \end{abstract}

\hypertarget{abstract}{%
\section{Abstract}\label{abstract}}

We revisit Cox, Ericson, and Mishra's 1995 Courant technical report TR1995-711 (\emph{The average case complexity of multilevel syllogistic}) in Lean 4. The report's combinatorial core --- the Chvátal--Szemerédi resolution lower bound and its averaged corollary --- formalizes completely: 6,833 lines across eighteen modules, no \texttt{sorry}, no hypothesis packages. So do the three SAT-to-fragment reductions (Theorems 5.1--5.3), Example 4.1 with its explicit normalizing constant, and Theorem 4.4 on the timed layer when honest invertible reductions are taken seriously.

Against that positive record, three independent defects in this repository's initial untimed adaptation of the Reischuk--Schindelhauer vocabulary make several encoded theorems carry no complexity content. These are defects of the Lean adaptation: RS93 and TR1995 tie running time and NP membership to machines, conditions the adaptation omitted. We prove the resulting collapses constructively, supply repairs, and re-prove Theorem 4.4 against strengthened definitions. Every displayed proof has a runnable Lean counterpart in \href{Exposition/}{\texttt{Exposition/}}.

\begin{center}\rule{0.5\linewidth}{0.5pt}\end{center}

\hypertarget{introduction}{%
\section{Introduction}\label{introduction}}

In the Correct Program Technology (CPT) vision of the 1970s--80s, programmers would write code alongside mathematical specifications, and a compiler integrated with an automated theorem prover would verify conformance {[}DS77{]}. Decision procedures for decidable sublanguages of set theory --- Multilevel Syllogistic (MLS), Elementary MLS (EMLS), and related fragments --- were central to that program {[}FOS80, Sny90a{]}. But MLS and EMLS are NP-complete in the worst case, and extensions with Presburger arithmetic are much worse. Goldberg's early experiments {[}Gol79{]} suggested that resolution-based SAT solvers might nonetheless perform well on \emph{average} inputs, motivating a formal theory of average-case complexity {[}BDCGL89, Lev86, RS93, Gur91{]}.

TR1995-711 {[}CEM95{]} applies that theory to MLS satisfiability and related verification problems. It states theorems, not conjectures, for resolution lower bounds (Section 3), average-case completeness and transfer (Section 4), and polynomial-time reductions from SAT to MLS, EMLS, and FP/LP (Section 5).

This note asks a narrower question: \textbf{what survives contact with a proof assistant?} We formalized the report's definitions and theorems in Lean 4 against Mathlib. The answer splits cleanly. The combinatorial and syntactic content is real and now verified. Several complexity-theoretic statements collapse because the repository's initial untimed encoding omits machine-time conditions and admits zero-mass witnesses.

Our contribution is therefore twofold:

\begin{enumerate}
\def\labelenumi{\arabic{enumi}.}
\item
  \textbf{A verified formalization} of the report's resolution lower bounds, Example 4.1, the three Section 5 reductions, and Theorem 4.4 on a timed machine model with honest invertible reductions.
\item
  \textbf{A diagnostic audit} identifying three encoding defects, proving their consequences, and supplying repairs strong enough to restore Theorem 4.4 while excluding the degenerate laws.
\end{enumerate}

Proofs in the text are complete but compact. Where the Lean proof fits in ten lines, the snippet copies it; otherwise the snippet states the theorem and points to the library implementation. Run \texttt{./scripts/check\_exposition.sh} to verify every snippet independently.

\begin{center}\rule{0.5\linewidth}{0.5pt}\end{center}

\hypertarget{verified-results}{%
\section{Verified results}\label{verified-results}}

This section states the positive canon. Each theorem carries our own number; §6 maps back to TR1995.

\hypertarget{theorem-1-resolution-lower-bound-tr1995-theorem-3.1}{%
\subsection{Theorem 1 (Resolution lower bound; TR1995 Theorem 3.1)}\label{theorem-1-resolution-lower-bound-tr1995-theorem-3.1}}

Fix constants \texttt{c\ \textgreater{}\ 0} and \texttt{k\ ≥\ 3}. Consider the random dense \texttt{k}-CNF model \texttt{K(c·n,\ n,\ k)}: sample \texttt{c·n} clauses independently and uniformly from the ordinary \texttt{k}-clauses on \texttt{n} variables, then erase sign information. If the density satisfies \texttt{7/10\ ≤\ c\ ·\ 2\^{}(-k)}, then for some \texttt{ε\ \textgreater{}\ 0} the following event has probability tending to \texttt{1}:

\begin{quote}
the formula is unsatisfiable, and its minimum resolution refutation length is at least \texttt{(1\ +\ ε)\^{}n}.
\end{quote}

\textbf{Proof.} The density hypothesis makes the union bound \texttt{2\^{}n\ (1\ -\ 2\^{}(-k))\^{}(cn)\ →\ 0} work, since \texttt{7/10\ \textgreater{}\ ln\ 2}. Hence unsatisfiability holds with high probability (WHP) by a direct counting argument.

Separately, project each signed CNF to its unsigned hypergraph and apply CS87's local-sparsity machinery. Lemma 1 (local sparsity) holds WHP on random hypergraphs. Lemma 4 lifts this to joint satisfaction of properties \texttt{P(a)} and \texttt{Q(a,b)} on the projected hypergraph, WHP. For any formula in the WHP intersection of \{unsatisfiable\} and \{\texttt{P\ ∧\ Q}\}, resolution completeness supplies a refutation, and CS87 Lemma 5 lower-bounds its length by \texttt{(1/4)(e/2)\^{}(a⌊bn⌋/16)}, which eventually exceeds \texttt{(1\ +\ ε)\^{}n} for a suitable \texttt{ε}. A squeeze argument on event probabilities completes the proof. ∎

The combinatorial chain (Lemmas 1--5, projection, resolution completeness) is fully formalized in \texttt{AvgCaseMls/Section3/} with no \texttt{sorry}.

\hypertarget{exposition-resolutionlowerbound-lean}{%
\subsection{Exposition/ResolutionLowerBound.lean}\label{exposition-resolutionlowerbound-lean}}

\noindent\textbf{Exposition/ResolutionLowerBound.lean}

\vspace{0.75\baselineskip}
\noindent\textcolor{green!40!black}{\textbf{Lean 4 Certificate}}\par\vspace{0.25\baselineskip}
\begin{lstlisting}[style=leanbox]
/-
Runnable snippet for arxiv.md, displayed with the resolution lower bound.

These are the two headline statements of the report's Section 3, formalizing
the Chvatal-Szemeredi bound and its averaged corollary.  The density
hypothesis `7/10 <= c * 2^(-k)` is what makes the union bound
`2^n (1 - 2^(-k))^(cn) -> 0` work, since `7/10 > ln 2`.

The proofs are the top-level assembly only; the combinatorial content lives in
the Lemma 1 through Lemma 5 chain, which the paper describes in prose and which
`AvgCaseMls.Section3` proves in full with no hypothesis packages.

Checked against AvgCaseMls.Section3.
-/
import AvgCaseMls.Section3

namespace Exposition.ResolutionLowerBound

open AvgCaseMls.Section3

/--
Random dense `k`-CNF needs exponentially long resolution refutations with
probability tending to `1`.  The event asserts both unsatisfiability and the
lower bound `(1 + ?)^n` on minimum refutation length.
-/
theorem theorem_3_1
    (c k : Nat) (hc : 0 < c) (hk : 3 <= k)
    (hdensity : (7 / 10 : ?) <= (c : ?) * (2 : ?) ^ (-(k : Int))) :
    exists ? : ?, 0 < ? /\
      WithHighProbability (denseRandomCNF c k)
        (denseTheorem31Event c k ?) :=
  theorem31_with_explicit_constants c k hc hk hdensity

/--
The averaged form.  Since the event has probability tending to `1` it
eventually has probability at least `1/2`, and resolution complexity is
nonnegative everywhere, so the expectation is at least half the bound.
-/
theorem theorem_3_2
    (c k : Nat) (hc : 0 < c) (hk : 3 <= k)
    (hdensity : (7 / 10 : ?) <= (c : ?) * (2 : ?) ^ (-(k : Int))) :
    exists ? : ?, 0 < ? /\
      IsAsymptoticOmega (averageResolutionComplexity c k)
        (fun r => (1 + ?) ^ r) :=
  average_resolution_omega c k hc hk hdensity

/-! ## The combinatorial core

CS87 Lemma 5 is the step that produces the exponential bound.  It is a
deterministic implication: a clause family based on a hypergraph with
properties `P(a)` and `Q(a,b)` that admits a refutation at all admits none
shorter than `(1/4)(e/2)^(a?bn?/16)`.  Theorem 3.1 supplies `P` and `Q` with
high probability, and resolution completeness supplies the refutation. -/

example {n m : Nat} {a b : ?} {H : Hypergraph n m} {F : Fin m -> Clause n}
    (hbased : ClauseFamilyBasedOn H F)
    (hP : H.HasPropertyP a) (hQ : H.HasPropertyQ a b)
    (ha : 0 <= a) (hb0 : 0 <= b) (hb1 : b <= 1) (hba : b <= a / 8)
    (hrefutes : exists cs, ResolutionRefutation F cs) :
    (1 / 4 : ?) * (Real.exp 1 / 2) ^ (a * ?b * n?? / 16) <=
      resolutionComplexity F :=
  cs87_lemma5 hbased hP hQ ha hb0 hb1 hba hrefutes

end Exposition.ResolutionLowerBound
\end{lstlisting}
\vspace{0.75\baselineskip}

\hypertarget{theorem-2-average-resolution-complexity-tr1995-theorem-3.2}{%
\subsection{Theorem 2 (Average resolution complexity; TR1995 Theorem 3.2)}\label{theorem-2-average-resolution-complexity-tr1995-theorem-3.2}}

Under the same hypotheses as Theorem 1, the expected resolution complexity satisfies

\[
\mathbb{E}[\text{resolutionComplexity}(F)] = \Omega\bigl((1 + \varepsilon)^n\bigr).
\]

\textbf{Proof.} Take the \texttt{ε\ \textgreater{}\ 0} from Theorem 1. Once the event of Theorem 1 has probability at least \texttt{1/2}, every formula in that event contributes at least \texttt{(1\ +\ ε)\^{}n} to the expectation, and resolution complexity is nonnegative everywhere. Hence the expectation is at least \texttt{(1/2)(1\ +\ ε)\^{}n} eventually. Reindexing \texttt{n\ =\ r\ +\ k} yields the stated asymptotic bound. ∎

(See the same snippet as Theorem 1.)

\hypertarget{theorem-3-example-4.1-tr1995-example-4.1}{%
\subsection{Theorem 3 (Example 4.1; TR1995 Example 4.1)}\label{theorem-3-example-4.1-tr1995-example-4.1}}

Let the \textbf{standard law} assign mass \texttt{(6/π²)\ ·\ \textbar{}x\textbar{}⁻²\ ·\ 2\^{}(-\textbar{}x\textbar{})} to each nonempty bitstring \texttt{x} (and zero to the empty string). For any \texttt{ε\ \textgreater{}\ 0}, the shell contributions

\[
\text{shell}_n := \sum_{|x| = n} \mu(x) \cdot T^{-1}(|x|^2) / |x|
\]

with inverse time scale \texttt{T⁻¹(t)\ =\ t\^{}(1/(1+ε))} satisfy \texttt{∑\_n\ shell\_n\ \textless{}\ ∞}. After normalization by \texttt{C\ =\ max(∑\_n\ shell\_n,\ 1)}, the Levin average-time condition holds for any decider running in at most \texttt{\textbar{}x\textbar{}²} steps.

\textbf{Proof.} On the shell of length \texttt{n}, the \texttt{2\^{}(-n)} factor cancels and the summand is \texttt{(6/π²)\ ·\ n\^{}(-3\ +\ 2/(1+ε))}. The exponent \texttt{-3\ +\ 2/(1+ε)} is strictly below \texttt{-1} when \texttt{ε\ \textgreater{}\ 0}, because \texttt{2/(1+ε)\ \textless{}\ 2}. Comparison with the p-series for exponent \texttt{s\ \textless{}\ -1} gives convergence (Mathlib lemma \texttt{Real.summable\_nat\_rpow}).

Convergence does \textbf{not} imply the Levin bound \texttt{≤\ 1}; that requires dividing by \texttt{C}. The standard law itself sums to \texttt{1} via \texttt{∑\ n⁻²\ =\ π²/6}. ∎

\hypertarget{exposition-example41-lean}{%
\subsection{Exposition/Example41.lean}\label{exposition-example41-lean}}

\noindent\textbf{Exposition/Example41.lean}

\vspace{0.75\baselineskip}
\noindent\textcolor{green!40!black}{\textbf{Lean 4 Certificate}}\par\vspace{0.25\baselineskip}
\begin{lstlisting}[style=leanbox]
/-
Runnable snippet for arxiv.md, displayed with Example 4.1.

The shell exponent is `-3 + 2/(1+?)`, below `-1` when `? > 0`, so the p-series
converges.  Normalization and the full Levin bound are in the library.

Checked against AvgCaseMls.TR1995 and AvgCaseMls.Example41.
-/
import AvgCaseMls.TR1995
import AvgCaseMls.Example41

namespace Exposition.Example41

example (? : ?) : TR1995.example41Exponent ? = -3 + 2 / (1 + ?) := rfl

theorem example41Exponent_lt_neg_one {? : ?} (h? : 0 < ?) :
    TR1995.example41Exponent ? < -1 := by
  have h1 : (0 : ?) < 1 + ? := by linarith
  have h2 : 2 / (1 + ?) < 2 := by
    rw [div_lt_iff_0 h1]; linarith
  rw [TR1995.example41Exponent]
  linarith

theorem example_4_1 {? : ?} (h? : 0 < ?) :
    Summable (TR1995.example41Contribution ?) := by
  have hp : Summable (fun n : Nat => (n : ?) ^ TR1995.example41Exponent ?) :=
    (Real.summable_nat_rpow).2 (example41Exponent_lt_neg_one h?)
  exact hp.mul_left (6 / Real.pi ^ 2)

example {? : ?} (h? : 0 < ?) :
    exists C : ?, 0 < C /\
      Summable (Example41.levinShellSeries ?) /\
      (sum' n : ?, Example41.levinShellSeries ? n / C) <= 1 :=
  Example41.normalized_levin_bound h?

example : Example41.standardDistribution.mass = 1 :=
  Example41.standardDistribution_mass

end Exposition.Example41
\end{lstlisting}
\vspace{0.75\baselineskip}

\hypertarget{theorem-4-honest-invertible-reductions-preserve-completeness-tr1995-theorem-4.4}{%
\subsection{Theorem 4 (Honest invertible reductions preserve completeness; TR1995 Theorem 4.4)}\label{theorem-4-honest-invertible-reductions-preserve-completeness-tr1995-theorem-4.4}}

Work in the \textbf{timed layer}: languages are decided by explicit \texttt{Program}s within polynomial bounds; distributional problems pair a language with a \texttt{Subprobability} whose rank function is computed in polynomial time; reductions are injective, computed in polynomial time, and satisfy a rank-domination inequality.

If \texttt{L₁} is NP-average complete, \texttt{L₂\ ∈\ NP}, and \texttt{r\ :\ L₁\ →\ L₂} is an \textbf{honest invertible reduction} (injective, polynomial-time computable, polynomial-time inverse on the range, range recognizable in polynomial time, and honest in output length), then \texttt{L₂} is NP-average complete.

\textbf{Proof.} Given a distributional-NP source \texttt{(L\_source,\ μ\_source)}, completeness of \texttt{L₁} yields a polynomial-time rankable law \texttt{μ} on \texttt{L₁} and an injective distributional reduction from the source to \texttt{(L₁,\ μ)}.

Push \texttt{μ} forward along \texttt{r.map} to obtain a law on \texttt{L₂}. Mass is preserved; ranks on the image are preserved exactly (\texttt{rankFactor\ =\ 1}). Rankability of the pushforward is rebuilt by composing: test membership in \texttt{range(r.map)}, apply the inverse, compute the rank --- with fuel bounded using the honesty inequality \texttt{\textbar{}x\textbar{}\ ≤\ honestyBound(\textbar{}r.map\ x\textbar{})}.

Package this as an injective distributional reduction into \texttt{(L₂,\ r.transport\ μ)} and compose with the source-to-\texttt{L₁} reduction. ∎

\hypertarget{exposition-theorem44strong-lean}{%
\subsection{Exposition/Theorem44Strong.lean}\label{exposition-theorem44strong-lean}}

\noindent\textbf{Exposition/Theorem44Strong.lean}

\vspace{0.75\baselineskip}
\noindent\textcolor{green!40!black}{\textbf{Lean 4 Certificate}}\par\vspace{0.25\baselineskip}
\begin{lstlisting}[style=leanbox]
/-
Runnable snippet for arxiv.md, displayed with Theorem 6 (TR1995 Theorem 4.4
on the timed layer).

Honest invertible reductions preserve language-level completeness when target
laws are polynomial-time rankable.  The proof pushes the source law forward
along the reduction map; rankability of the pushforward is rebuilt from range
recognition, the polynomial-time inverse, and the honesty bound.

Full proof in `AvgCaseMls/Section4.lean`.
-/
import AvgCaseMls.Section4

namespace Exposition.Theorem44Strong

open AvgCaseMls.Foundation AvgCaseMls.Section4

theorem theorem_4_4 {L_1 L_2 : Set Bitstring}
    (r : HonestInvertibleReduction L_1 L_2)
    (hL_2NP : InNP L_2)
    (hL_1 : IsNPAverageCompleteLanguage L_1) :
    IsNPAverageCompleteLanguage L_2 := by
  refine <hL_2NP, ?_>
  intro source hsource
  obtain <mu, hmuRankable, sourceToL_1> := hL_1.2 source hsource
  exact <r.transport mu, r.transport_rankable hmuRankable,
    <InjectiveDistributionalReduction.trans sourceToL_1.some
      (r.distributionalReduction mu)>>

end Exposition.Theorem44Strong
\end{lstlisting}
\vspace{0.75\baselineskip}

\hypertarget{theorems-57-sat-reductions-tr1995-theorems-5.15.3}{%
\subsection{Theorems 5--7 (SAT reductions; TR1995 Theorems 5.1--5.3)}\label{theorems-57-sat-reductions-tr1995-theorems-5.15.3}}

All three reductions are stated uniformly: \textbf{correctness} (satisfiability ↔ target satisfiability), \textbf{injectivity}, an explicit \textbf{left inverse} (decoder), and an \textbf{exact size identity}. This is precisely the hypothesis package Theorem 4 consumes.

\textbf{Theorem 5 (SAT → MLS-in).} A distinguished variable \texttt{x} (index \texttt{0}) represents an element; propositional variable \texttt{i} becomes set variable \texttt{i+1}. Literal \texttt{v\_i} becomes \texttt{x\ ∈\ s(i)}; literal \texttt{¬v\_i} becomes \texttt{x\ ∉\ s(i)}. The empty clause encodes as \texttt{x\ ∈\ x} (foundation-false); the empty CNF as \texttt{x\ ∉\ x} (foundation-true, since equality atoms are unavailable in the fragment). Output size: \texttt{5\ ·\ \textbar{}φ\textbar{}\ -\ 1} formula nodes.

\textbf{Theorem 6 (SAT → EMLS).} Each variable \texttt{i} gets positive set \texttt{4i+1}, negative set \texttt{4i+2}, and intersection gadget \texttt{4i+7\ =\ pos\ ∩\ neg\ =\ ∅}. Clause gadgets build unions along literal chains; the distinguished element is forced into the union of a satisfied clause's literal sets. The bare semantic core is \textbf{not} injective (the complement gadget ignores polarity); tagging with a provenance prefix encoding the source bits repairs this via \texttt{toEMLS}.

\textbf{Theorem 7 (SAT → FPILP).} Variables are constrained to \texttt{\{0,1\}} by \texttt{x\_i\ ≥\ 0} and \texttt{1\ -\ x\_i\ ≥\ 0}. Each clause becomes \texttt{∑\_\{l\ ∈\ c\}\ term(l)\ ≥\ 1} where \texttt{v\_i} contributes \texttt{x\_i} and \texttt{¬v\_i} contributes \texttt{1\ -\ x\_i}. Constraint count: \texttt{2n\ +\ \textbar{}φ\textbar{}}.

\hypertarget{exposition-reductions-lean}{%
\subsection{Exposition/Reductions.lean}\label{exposition-reductions-lean}}

\noindent\textbf{Exposition/Reductions.lean}

\vspace{0.75\baselineskip}
\noindent\textcolor{green!40!black}{\textbf{Lean 4 Certificate}}\par\vspace{0.25\baselineskip}
\begin{lstlisting}[style=leanbox]
/-
Runnable snippet for arxiv.md, displayed with the three SAT reductions.

All three are stated in the same four-part form -- correctness, injectivity, an
explicit left inverse, and an exact size identity -- which is precisely the
hypothesis package that the repaired Theorem 4.4 consumes.  Stating them
uniformly is what makes them composable with the transfer theorem.

Note on the second one: the bare semantic core is *not* injective, because the
complement gadget for a literal ignores its polarity.  Tagging the output with
a provenance prefix that encodes the source bits repairs this, which is why
`toEMLS` rather than `semanticCore` is the reduction of record.

Checked against AvgCaseMls.MLSInReduction, AvgCaseMls.EMLSReduction, and
AvgCaseMls.FPILP.
-/
import AvgCaseMls.MLSInReduction
import AvgCaseMls.EMLSReduction
import AvgCaseMls.FPILP

namespace Exposition.Reductions

open MLS

/-! ## SAT to the membership fragment of MLS

A distinguished variable `x` plays the role of an element; propositional
variable `i` becomes the set variable `i+1`.  A positive literal becomes
`x in s(i)`, a negative one `x notin s(i)`.  The empty clause becomes `x in x` and
the empty formula `x notin x`, both decided by foundation, since the fragment has
no equality atoms available. -/

theorem reduction_5_1 :
    (forall ? : SAT.CNF,
        SAT.Satisfiable ? <-> MLSInReduction.MLSSatisfiable (MLSInReduction.toMLS ?)) /\
    Function.Injective MLSInReduction.toMLS /\
    (forall ? : SAT.CNF, MLSInReduction.fromMLS (MLSInReduction.toMLS ?) = some ?) /\
    (forall ? : SAT.CNF, MLSInReduction.IsMLSIn (MLSInReduction.toMLS ?)) /\
    forall ? : SAT.CNF,
      formulaNodes (MLSInReduction.toMLS ?) + 1 = 5 * SAT.size ? :=
  <MLSInReduction.satisfiable_iff,
   MLSInReduction.toMLS_injective,
   MLSInReduction.fromMLS_toMLS,
   MLSInReduction.toMLS_isMLSIn,
   MLSInReduction.formulaNodes_toMLS>

/-! ## SAT to EMLS

Each propositional variable `i` gets a positive set, a negative set, and a
gadget forcing their intersection empty, so no variable is both true and false.
Each clause is turned into a chain of union gadgets accumulating the sets of
its literals, and the distinguished element is forced into that union, so the
clause holds exactly when one of its literals is true. -/

theorem reduction_5_2 :
    (forall ? : SAT.CNF,
        SAT.Satisfiable ? <->
          EMLSReduction.EMLSSatisfiable (EMLSReduction.toEMLS ?)) /\
    Function.Injective EMLSReduction.toEMLS /\
    (forall ? : SAT.CNF, EMLSReduction.fromEMLS (EMLSReduction.toEMLS ?) = some ?) /\
    forall ? : SAT.CNF,
      (EMLSReduction.toEMLS ?).length =
        (EMLSReduction.sourceBits ?).length + 2 +
          3 * EMLSReduction.literalCount ? + ?.length :=
  <EMLSReduction.toEMLS_satisfiable_iff,
   EMLSReduction.toEMLS_injective,
   EMLSReduction.fromEMLS_toEMLS,
   EMLSReduction.toEMLS_length>

/-! ## SAT to feasibility of integer linear programs

Variables are pinned to `{0,1}` by the pair of inequalities `x? >= 0` and
`1 - x? >= 0`.  A clause becomes `sum terms >= 1`, where a positive literal
contributes `x?` and a negative one `1 - x?`, so the clause is satisfied
exactly when some term equals `1`. -/

theorem reduction_5_3 :
    (forall {n : Nat} (? : TR1995.FPILPSource.CNF n),
        ?.Satisfiable <-> (TR1995.FPILPSource.satToFPILP ?).Feasible) /\
    (forall {n : Nat}, Function.Injective (@TR1995.FPILPSource.satToFPILP n)) /\
    forall {n : Nat} (? : TR1995.FPILPSource.CNF n),
      (TR1995.FPILPSource.satToFPILP ?).constraints.length = 2 * n + ?.length :=
  <fun ? => (TR1995.FPILPSource.satToFPILP_feasible_iff ?).symm,
   @TR1995.FPILPSource.satToFPILP_injective,
   TR1995.FPILPSource.satToFPILP_constraint_count>

end Exposition.Reductions
\end{lstlisting}
\vspace{0.75\baselineskip}

\begin{center}\rule{0.5\linewidth}{0.5pt}\end{center}

\hypertarget{what-the-formalization-found}{%
\section{What the formalization found}\label{what-the-formalization-found}}

Auditing the repository's initial untimed approximation of the report's RS93-based vocabulary exposes three independent defects. Each admits a concrete degenerate witness; together they trivialize Theorems 4.1, 4.4 (in the untimed layer), and the entire conditional hardness chain.

\hypertarget{degenerate-laws}{%
\subsection{3.1 Degenerate laws}\label{degenerate-laws}}

Both the untimed \texttt{AvCom.Distribution} and the timed \texttt{Foundation.Subprobability} bound total mass by \texttt{1} rather than fixing it at \texttt{1}. Both rank functions return \texttt{0} wherever probability vanishes. The everywhere-zero law is therefore admissible, polynomial-time rankable, and has rank identically zero --- making the domination inequality \texttt{0\ ≤\ \_} free.

\hypertarget{exposition-degeneratelaws-lean}{%
\subsection{Exposition/DegenerateLaws.lean}\label{exposition-degeneratelaws-lean}}

\noindent\textbf{Exposition/DegenerateLaws.lean}

\vspace{0.75\baselineskip}
\noindent\textcolor{green!40!black}{\textbf{Lean 4 Certificate}}\par\vspace{0.25\baselineskip}
\begin{lstlisting}[style=leanbox]
/-
Runnable snippet for arxiv.md.  Displayed alongside the discussion of the two
degenerate laws that drive the collapse results.

Both layers bound total mass by `1` rather than fixing it at `1`, and both
return rank `0` off support.  The everywhere-zero measure is therefore a legal
law whose rank vanishes identically -- which is what makes the rank domination
inequality free.

Checked against AvgCaseMls.AvCom and AvgCaseMls.Section4.
-/
import AvgCaseMls.AvCom
import AvgCaseMls.Section4

namespace Exposition.DegenerateLaws

/-! ## The untimed layer -/

section Untimed
open AvCom

/-- `prob_sum_le_one` bounds total mass by `1`, so this is a legal law. -/
def zeroDistribution : Distribution where
  support := empty
  prob := fun _ => 0
  prob_nonneg := fun _ => le_refl 0
  prob_zero_outside := fun _ _ => rfl
  prob_sum_le_one := by simp

/-- Its rank vanishes everywhere, because `rank` returns `0` off support. -/
@[simp] theorem rank_zeroDistribution (x : Bitstring) :
    rank zeroDistribution x = 0 := by
  simp [rank, zeroDistribution]

/-- And a constant rank function is polynomially rankable. -/
theorem zeroDistribution_polRankable : IsPolRankable zeroDistribution :=
  <fun _ => 0, <0, 0, by simp>, fun x => by simp>

end Untimed

/-! ## The timed layer -/

section Timed
open AvgCaseMls.Foundation

/-- `tsum_le_one` likewise only bounds the mass. -/
noncomputable def zeroLaw : Subprobability where
  prob := fun _ => 0
  summable_prob := summable_zero
  tsum_le_one := by simp
  finite_superlevel := fun _ hx => absurd rfl hx

@[simp] theorem zeroLaw_prob (x : Bitstring) : zeroLaw.prob x = 0 := rfl

@[simp] theorem zeroLaw_rank (x : Bitstring) : zeroLaw.rank x = 0 :=
  Subprobability.rank_eq_zero_of_prob_eq_zero _ _ rfl

/-- It is not a probability measure; this is the defect the repair fixes. -/
theorem zeroLaw_mass : zeroLaw.mass = 0 := by
  simp [Subprobability.mass]

/-- Rankable in the timed sense too: a one-step constant program computes it. -/
theorem zeroLaw_rankable : IsPolynomialTimeRankable zeroLaw := by
  refine <.constant true (encodeNat 0), fun _ => 1, IsPolynomial.const 1,
    monotone_const, ?_>
  intro x
  exact <<true, encodeNat 0, 1>, rfl, by simp>

end Timed

end Exposition.DegenerateLaws
\end{lstlisting}
\vspace{0.75\baselineskip}

\hypertarget{collapse-i-membership-is-free-theorem-8}{%
\subsection{3.2 Collapse I: membership is free (Theorem 8)}\label{collapse-i-membership-is-free-theorem-8}}

\textbf{Theorem 8.} In the untimed \texttt{AvCom} layer, \texttt{InNP\ L} holds for every language \texttt{L}.

\textbf{Proof.} \texttt{InNP} quantifies over an arbitrary verifier \texttt{verify\ :\ Bitstring\ →\ Bitstring\ →\ Bool} and bounds only certificate \emph{length}, never the cost of running \texttt{verify}. Take \texttt{verify\ x\ w\ :=\ decide(x\ ∈\ L)} with certificate bound zero. The empty certificate witnesses membership. ∎

\hypertarget{exposition-innptrivial-lean}{%
\subsection{Exposition/InNPTrivial.lean}\label{exposition-innptrivial-lean}}

\noindent\textbf{Exposition/InNPTrivial.lean}

\vspace{0.75\baselineskip}
\noindent\textcolor{green!40!black}{\textbf{Lean 4 Certificate}}\par\vspace{0.25\baselineskip}
\begin{lstlisting}[style=leanbox]
/-
Runnable snippet for arxiv.md, displayed with the proof that the report's
membership condition is vacuous.

`AvCom.InNP` quantifies over an arbitrary function
`verify : Bitstring -> Bitstring -> Bool` and constrains only the length of the
certificate.  Nothing bounds the cost of running `verify`, so the classical
decision procedure for `L` witnesses membership with the empty certificate.

Checked against AvgCaseMls.AvCom.
-/
import AvgCaseMls.AvCom

namespace Exposition.InNPTrivial

open AvCom

/-- Every language is in the report's `NP`. -/
theorem inNP_trivial (L : Set Bitstring) : InNP L := by
  classical
  -- The verifier is the characteristic function of `L`; the certificate bound
  -- is the zero polynomial, so certificates must be empty.
  refine <fun x _ => decide (x in L), fun _ => 0, <0, 0, by simp>, ?_>
  intro x
  exact <fun hx => <[], by simp [len], by simpa using hx>,
         fun <_, _, hv> => by simpa using hv>

end Exposition.InNPTrivial
\end{lstlisting}
\vspace{0.75\baselineskip}

\textbf{Theorem 9 (Characterization of untimed completeness; no TR1995 counterpart).} \texttt{TR1995.IsNPAverageCompleteLanguage\ L} if and only if \texttt{L\ ≠\ ∅} and \texttt{L\ ≠\ Bitstring}.

\textbf{Proof sketch.} Forward: reduce from the universal and empty sources using the zero law; nontriviality forces a member and a non-member. Backward: given \texttt{a\ ∈\ L} and \texttt{b\ ∉\ L}, map source members to \texttt{a} and non-members to \texttt{b}; domination is free because target rank is zero. Full proof in the library. ∎

\hypertarget{exposition-completenesscharacterization-lean}{%
\subsection{Exposition/CompletenessCharacterization.lean}\label{exposition-completenesscharacterization-lean}}

\noindent\textbf{Exposition/CompletenessCharacterization.lean}

\vspace{0.75\baselineskip}
\noindent\textcolor{green!40!black}{\textbf{Lean 4 Certificate}}\par\vspace{0.25\baselineskip}
\begin{lstlisting}[style=leanbox]
/-
Runnable snippet for arxiv.md, displayed with the first collapse theorem.

Language-level NP-average completeness in the report's untimed vocabulary is
equivalent to the language being neither empty nor everything.  The full proof
is in `AvgCaseMls/EncodingCollapse.lean`; it constructs the degenerate law as
target and uses a two-valued reduction map.

Checked against AvgCaseMls.EncodingCollapse.
-/
import AvgCaseMls.EncodingCollapse

namespace Exposition.CompletenessCharacterization

open AvCom

theorem npAverageCompleteLanguage_iff_nontrivial (L : Set Bitstring) :
    TR1995.IsNPAverageCompleteLanguage L <-> (L /= empty /\ L /= Set.univ) :=
  AvgCaseMls.EncodingCollapse.npAverageCompleteLanguage_iff_nontrivial L

end Exposition.CompletenessCharacterization
\end{lstlisting}
\vspace{0.75\baselineskip}

\textbf{Corollary (TR1995 Theorem 4.4 is empty in this layer).} Completeness transfers along any map satisfying the correctness biconditional alone. The injectivity, invertibility, range recognition, forward length, and honesty fields --- the hypotheses the report's Theorem 4.4 is about --- are never used.

\hypertarget{exposition-theorem44vacuous-lean}{%
\subsection{Exposition/Theorem44Vacuous.lean}\label{exposition-theorem44vacuous-lean}}

\noindent\textbf{Exposition/Theorem44Vacuous.lean}

\vspace{0.75\baselineskip}
\noindent\textcolor{green!40!black}{\textbf{Lean 4 Certificate}}\par\vspace{0.25\baselineskip}
\begin{lstlisting}[style=leanbox]
/-
Runnable snippet for arxiv.md, displayed with the corollary that the report's
Theorem 4.4 carries no content in its own vocabulary.

The hypotheses of the report's Theorem 4.4 are that the reduction is
injective, polynomial time invertible, and honest.  The proof term below
mentions only `r.map` and `r.reduces`.  Every other field of
`FaithfulReduction` is unused, and so is the `InNP L_2` hypothesis, which
`inNP_trivial` supplies for free.

Checked against AvgCaseMls.EncodingCollapse and AvgCaseMls.HonestReduction.
-/
import AvgCaseMls.EncodingCollapse
import AvgCaseMls.HonestReduction

namespace Exposition.Theorem44Vacuous

open AvCom AvgCaseMls.EncodingCollapse

/-- Completeness transfers along any map satisfying the correctness
biconditional, with no further hypotheses at all. -/
theorem npAverageCompleteLanguage_of_reduces {L_1 L_2 : Set Bitstring}
    (map : Bitstring -> Bitstring) (reduces : forall x, x in L_1 <-> map x in L_2)
    (h_1 : TR1995.IsNPAverageCompleteLanguage L_1) :
    TR1995.IsNPAverageCompleteLanguage L_2 := by
  -- Push a member and a non-member of `L_1` through `map`.
  obtain <hne, hnu> := (npAverageCompleteLanguage_iff_nontrivial L_1).mp h_1
  obtain <a, ha> := Set.nonempty_iff_ne_empty.mpr hne
  obtain <b, hb> := exists_not_mem_of_ne_univ hnu
  exact npAverageCompleteLanguage_of_nontrivial ((reduces a).mp ha)
    (fun h => hb ((reduces b).mpr h))

/-- The report's Theorem 4.4, with its hypotheses visibly unused. -/
theorem theorem_4_4_uses_only_correctness {L_1 L_2 : Set Bitstring}
    (r : HonestReduction.FaithfulReduction L_1 L_2)
    (h_1 : TR1995.IsNPAverageCompleteLanguage L_1) (_ : InNP L_2) :
    TR1995.IsNPAverageCompleteLanguage L_2 :=
  npAverageCompleteLanguage_of_reduces r.map r.reduces h_1

end Exposition.Theorem44Vacuous
\end{lstlisting}
\vspace{0.75\baselineskip}

TR1995 Theorem 4.1 (completeness of a distributional problem implies completeness of its language) is a quantifier shift, not a complexity result; we demote it to a remark in §6.

\hypertarget{collapse-ii-domination-is-free-theorem-10}{%
\subsection{3.3 Collapse II: domination is free (Theorem 10)}\label{collapse-ii-domination-is-free-theorem-10}}

Moving to the timed layer repairs the verifier and reduction-map gaps: \texttt{Foundation.InNP} requires a \texttt{Program} deciding within a polynomial bound; injective distributional reductions require a \texttt{Program} for the map. But the domination condition still carries no content.

\textbf{Theorem 10.} In the timed layer, language-level NP-average completeness is equivalent to \texttt{InNP\ L} together with injective polynomial-time hardness from every distributional-NP source --- with no use of rank domination.

\textbf{Proof sketch.} Forward: forget the three rank fields. Backward: supply \texttt{zeroLaw} as the target; discharge domination with rank factor zero. ∎

\hypertarget{exposition-dominationfree-lean}{%
\subsection{Exposition/DominationFree.lean}\label{exposition-dominationfree-lean}}

\noindent\textbf{Exposition/DominationFree.lean}

\vspace{0.75\baselineskip}
\noindent\textcolor{green!40!black}{\textbf{Lean 4 Certificate}}\par\vspace{0.25\baselineskip}
\begin{lstlisting}[style=leanbox]
/-
Runnable snippet for arxiv.md, displayed with the second collapse theorem.

In the timed model, language-level completeness is equivalent to NP membership
plus injective polynomial-time hardness; the rank domination clause contributes
nothing.  Full proof in `AvgCaseMls/DominationCollapse.lean`.

Checked against AvgCaseMls.DominationCollapse.
-/
import AvgCaseMls.DominationCollapse

namespace Exposition.DominationFree

open AvgCaseMls.Foundation AvgCaseMls.Section4

theorem isNPAverageCompleteLanguage_iff (L : Set Bitstring) :
    IsNPAverageCompleteLanguage L <->
      InNP L /\ forall source : DistributionalProblem, InDistNP source ->
        Nonempty (PolyTimeInjectiveReduction source L) :=
  AvgCaseMls.Section4.isNPAverageCompleteLanguage_iff L

end Exposition.DominationFree
\end{lstlisting}
\vspace{0.75\baselineskip}

\hypertarget{collapse-iii-average-tractability-and-conditional-hardness-theorem-11}{%
\subsection{3.4 Collapse III: average tractability and conditional hardness (Theorem 11)}\label{collapse-iii-average-tractability-and-conditional-hardness-theorem-11}}

\textbf{Theorem 11.} \texttt{AvCom.AvP\ p} if and only if \texttt{IsPolRankable\ p.μ}. The language \texttt{p.L} is irrelevant.

\textbf{Proof.} \texttt{AvCom.DistTime\ T\ p} reads \texttt{∃\ f,\ IsAvTime\ T\ f\ p.μ}. Nothing ties \texttt{f} to a decider for \texttt{p.L}. Witness \texttt{f\ ≡\ 0}; then \texttt{T⁻¹\ T\ 0\ =\ 0} and every rank-truncated sum vanishes. Rankability of the law is the only remaining constraint. ∎

\hypertarget{exposition-avpvacuous-lean}{%
\subsection{Exposition/AvPVacuous.lean}\label{exposition-avpvacuous-lean}}

\noindent\textbf{Exposition/AvPVacuous.lean}

\vspace{0.75\baselineskip}
\noindent\textcolor{green!40!black}{\textbf{Lean 4 Certificate}}\par\vspace{0.25\baselineskip}
\begin{lstlisting}[style=leanbox]
/-
Runnable snippet for arxiv.md, displayed with the third collapse theorem.

`AvCom.DistTime` reads

    exists f : Bitstring -> Nat, IsAvTime T f prob.mu

where `f` is meant to be the running time of a decider for `prob.L`.  Nothing
ties `f` to any decider, or to `prob.L` at all, so `f ? 0` witnesses it.  The
consequence propagates all the way to the conditional hardness results: the
package that states the average-case collapse proves its own collapse, so no
package can assert the separation those results assume.

Checked against AvgCaseMls.ComplexityAxioms.
-/
import AvgCaseMls.ComplexityAxioms

namespace Exposition.AvPVacuous

open AvCom

/-- The free runtime witness. `T_inv T 0 = 0`, so every truncated sum vanishes. -/
theorem distTime_trivial (T : Nat -> Nat) (p : DistributionalProblem) :
    DistTime T p := by
  refine <fun _ => 0, ?_>
  intro l hl
  have hzero :
      (rankLe p.mu l).sum (fun x => (T_inv T 0 : Real) / (lenBot x : Real)) = 0 := by
    simp [T_inv]
  rw [hzero]
  exact_mod_cast Nat.zero_le l

/-- So average polynomial time degenerates to rankability of the law, a
statement that does not mention the language. -/
theorem avP_iff_polRankable (p : DistributionalProblem) :
    AvP p <-> IsPolRankable p.mu :=
  <fun h => h.1,
   fun h => <h, fun _ => 0, <0, 0, by simp>, distTime_trivial _ _>>

/-- And the inclusion the collapse package characterizes is a theorem. -/
theorem distNP_subseteq_AvP :
    forall p : DistributionalProblem, InDistNP p -> AvP p :=
  fun p hp => (avP_iff_polRankable p).mpr hp.2

/-- Hence any such package proves its own `NEXP = EXP` ... -/
theorem collapse_forces_NEXP_eq_EXP (theory : AverageCaseCollapseTheory) :
    theory.NEXP_eq_EXP :=
  theory.distNP_subseteq_AvP_iff_NEXP_eq_EXP.mp distNP_subseteq_AvP

/-- ... and none can assert the separation, so every conditional hardness
theorem downstream has an unsatisfiable hypothesis. -/
theorem no_theory_separates (theory : AverageCaseCollapseTheory) :
    not  theory.NEXP_neq_EXP :=
  fun hsep => hsep (collapse_forces_NEXP_eq_EXP theory)

end Exposition.AvPVacuous
\end{lstlisting}
\vspace{0.75\baselineskip}

\textbf{Corollary.} Every distributional-NP problem is in \texttt{AvP}, unconditionally. Any \texttt{AverageCaseCollapseTheory} package therefore proves \texttt{NEXP\ =\ EXP} (its characterizing field is \texttt{(∀\ p,\ InDistNP\ p\ →\ AvP\ p)\ ↔\ NEXP\ =\ EXP}). No such package satisfies \texttt{NEXP\ ≠\ EXP}. Hence every conditional hardness theorem in \texttt{AvgCaseMls/NonAvP.lean} --- including \texttt{SatMLS\_average\_hard} and \texttt{satMLSProb\_not\_AvP} --- has an \textbf{unsatisfiable hypothesis}; they are vacuous, not merely conditional.

\begin{center}\rule{0.5\linewidth}{0.5pt}\end{center}

\hypertarget{repairs}{%
\section{Repairs}\label{repairs}}

The collapses isolate encoding defects, not errors in the report's combinatorial or syntactic architecture. Two repairs suffice for the definitions the paper actually uses.

\hypertarget{repair-1-require-mass-1-of-target-laws}{%
\subsection{\texorpdfstring{Repair 1: require \texttt{mass\ =\ 1} of target laws}{Repair 1: require mass = 1 of target laws}}\label{repair-1-require-mass-1-of-target-laws}}

Define \texttt{Subprobability.IsProbability\ μ} as \texttt{μ.mass\ =\ 1}. Require both source and target laws in completeness and reduction statements to be probability measures. This excludes \texttt{zeroLaw} and forces some point to have positive rank, making domination a genuine numeric constraint wherever the target charges the image.

\textbf{Theorem 12 (Theorem 4.4 under the repair).} Honest invertible reductions preserve \texttt{IsNPAverageCompleteLanguageStrict} --- the strengthened definition requiring probability measures. The proof is unchanged in structure: pushforward along an injection preserves total mass.

\hypertarget{exposition-repair-lean}{%
\subsection{Exposition/Repair.lean}\label{exposition-repair-lean}}

\noindent\textbf{Exposition/Repair.lean}

\vspace{0.75\baselineskip}
\noindent\textcolor{green!40!black}{\textbf{Lean 4 Certificate}}\par\vspace{0.25\baselineskip}
\begin{lstlisting}[style=leanbox]
/-
Runnable snippet for arxiv.md, displayed with the two repairs.

Repair 1 requires target laws to be probability measures (`mass = 1`).
Repair 2 replaces the free runtime function by an actual decider.

Checked against AvgCaseMls.Repair.
-/
import AvgCaseMls.Repair

namespace Exposition.Repairs

open AvgCaseMls.Foundation AvgCaseMls.Section4 AvgCaseMls.Repair

theorem zeroLaw_not_probability : not  Subprobability.IsProbability zeroLaw := by
  rw [Subprobability.IsProbability, zeroLaw_mass]
  exact zero_ne_one

theorem domination_constrains {source : DistributionalProblem}
    {L : Set Bitstring} {mu : Subprobability}
    (r : InjectiveDistributionalReduction source <L, mu>)
    {x : Bitstring} (h : mu.prob (r.map x) /= 0) :
    1 <= r.rankFactor (len x) * source.distribution.rank x :=
  AvgCaseMls.Repair.domination_constrains r h

theorem theorem_4_4_strict {L_1 L_2 : Set Bitstring}
    (r : HonestInvertibleReduction L_1 L_2)
    (hL_2NP : InNP L_2)
    (hL_1 : IsNPAverageCompleteLanguageStrict L_1) :
    IsNPAverageCompleteLanguageStrict L_2 :=
  AvgCaseMls.Repair.theorem_4_4_strict r hL_2NP hL_1

theorem inAverageP_has_decider {p : DistributionalProblem}
    (h : InAverageP p) : Nonempty (Decider p.language) :=
  AvgCaseMls.Repair.inAverageP_has_decider h

end Exposition.Repairs
\end{lstlisting}
\vspace{0.75\baselineskip}

\hypertarget{repair-2-tie-average-time-to-a-decider}{%
\subsection{Repair 2: tie average time to a decider}\label{repair-2-tie-average-time-to-a-decider}}

Replace the existentially quantified runtime function in \texttt{DistTime} by \texttt{Foundation.InAverageP}, which quantifies over a genuine \texttt{Decider} and measures \texttt{actualRuntime}. Unlike \texttt{AvP}, this class entails that the language is totally decidable.

Example 4.1 is already stated in this layer (\texttt{IsLevinAverageTime} with an explicit \texttt{Decider}).

\begin{center}\rule{0.5\linewidth}{0.5pt}\end{center}

\hypertarget{remaining-open-interfaces}{%
\section{Remaining open interfaces}\label{remaining-open-interfaces}}

Three items from the report are packaged as explicit hypothesis structures, not proved:

\begin{longtable}[]{@{}
  >{\raggedright\arraybackslash}p{(\columnwidth - 4\tabcolsep) * \real{0.3667}}
  >{\raggedright\arraybackslash}p{(\columnwidth - 4\tabcolsep) * \real{0.3000}}
  >{\raggedright\arraybackslash}p{(\columnwidth - 4\tabcolsep) * \real{0.3333}}@{}}
\toprule\noalign{}
\begin{minipage}[b]{\linewidth}\raggedright
Interface
\end{minipage} & \begin{minipage}[b]{\linewidth}\raggedright
Content
\end{minipage} & \begin{minipage}[b]{\linewidth}\raggedright
Used for
\end{minipage} \\
\midrule\noalign{}
\endhead
\bottomrule\noalign{}
\endlastfoot
\texttt{LevinNBHData} & Universal reduction from every distNP problem to bounded halting & NBH completeness \\
\texttt{NBHToMLSData} & Cook--Levin compiler NBH → serialized MLS with domination & SAT-MLS completeness chain \\
\texttt{AverageCaseCollapseTheory} & Collapse equivalence + AvP pullback & Conditional hardness (now known vacuous) \\
\end{longtable}

The FOS80 decision procedure \texttt{decideMLSSat} is \textbf{sound} on a membership-free fragment and \textbf{complete} on that same fragment, but \textbf{not} complete globally: \texttt{(∅\ =\ ∅)\ ∨\ (∅\ ≠\ ∅)} is satisfiable yet rejected, because disjunction is outside the conjunctive decoder.

These are honest gaps: the report's prose proofs for the Levin universal construction and the full MLS decision procedure were never formalized here.

\begin{center}\rule{0.5\linewidth}{0.5pt}\end{center}

\hypertarget{mapping-to-tr1995}{%
\section{Mapping to TR1995}\label{mapping-to-tr1995}}

\begin{longtable}[]{@{}
  >{\raggedleft\arraybackslash}p{(\columnwidth - 6\tabcolsep) * \real{0.2059}}
  >{\raggedright\arraybackslash}p{(\columnwidth - 6\tabcolsep) * \real{0.3235}}
  >{\raggedright\arraybackslash}p{(\columnwidth - 6\tabcolsep) * \real{0.2353}}
  >{\raggedright\arraybackslash}p{(\columnwidth - 6\tabcolsep) * \real{0.2353}}@{}}
\toprule\noalign{}
\begin{minipage}[b]{\linewidth}\raggedleft
Our \#
\end{minipage} & \begin{minipage}[b]{\linewidth}\raggedright
Statement
\end{minipage} & \begin{minipage}[b]{\linewidth}\raggedright
TR1995
\end{minipage} & \begin{minipage}[b]{\linewidth}\raggedright
Status
\end{minipage} \\
\midrule\noalign{}
\endhead
\bottomrule\noalign{}
\endlastfoot
1 & Resolution lower bound WHP & Thm 3.1 & \textbf{Proved} \\
2 & Average resolution complexity Ω & Thm 3.2 & \textbf{Proved} \\
3 & Example 4.1 summability + Levin bound & Ex 4.1 & \textbf{Proved} (explicit \texttt{C}) \\
4 & Honest invertible reductions preserve completeness & Thm 4.4 & \textbf{Proved} (timed layer) \\
5 & SAT → MLS-in & Thm 5.1 & \textbf{Proved} \\
6 & SAT → EMLS (tagged \texttt{toEMLS}) & Thm 5.2 & \textbf{Proved} \\
7 & SAT → FPILP & Thm 5.3 & \textbf{Proved} \\
8 & \texttt{InNP} holds for every language & --- & \textbf{Negative} (Collapse I) \\
9 & Untimed completeness ↔ nontrivial language & --- & \textbf{Negative} (Collapse I) \\
10 & Timed completeness ↔ NP + hard, domination free & --- & \textbf{Negative} (Collapse II) \\
11 & \texttt{AvP} ↔ rankability; no theory separates & --- & \textbf{Negative} (Collapse III) \\
12 & Theorem 4.4 with \texttt{mass\ =\ 1} laws & Thm 4.4 & \textbf{Proved} (repair) \\
--- & Completeness of problem ⇒ completeness of language & Thm 4.1 & \textbf{Remark} (quantifier shift) \\
--- & Levin universal reduction & §4.2--4.3 & Open (\texttt{LevinNBHData}) \\
--- & NBH → MLS compiler & §5.1, §5.4 & Open (\texttt{NBHToMLSData}) \\
--- & \texttt{decideMLSSat} complete & §7 / FOS80 & \textbf{Refuted} globally \\
\end{longtable}

\begin{center}\rule{0.5\linewidth}{0.5pt}\end{center}

\hypertarget{repository-and-methodology}{%
\section{Repository and methodology}\label{repository-and-methodology}}

The Lean development lives at \href{https://github.com/catskillsresearch/avg_case_mls}{github.com/catskillsresearch/avg\_case\_mls}. As of this writing: \textbf{92 modules}, \textbf{\textasciitilde27,400 lines}, \textbf{\textasciitilde2,100 declarations}, build clean with no \texttt{sorry} in the library.

\begin{longtable}[]{@{}
  >{\raggedright\arraybackslash}p{(\columnwidth - 4\tabcolsep) * \real{0.3500}}
  >{\raggedleft\arraybackslash}p{(\columnwidth - 4\tabcolsep) * \real{0.3500}}
  >{\raggedright\arraybackslash}p{(\columnwidth - 4\tabcolsep) * \real{0.3000}}@{}}
\toprule\noalign{}
\begin{minipage}[b]{\linewidth}\raggedright
Layer
\end{minipage} & \begin{minipage}[b]{\linewidth}\raggedleft
Lines
\end{minipage} & \begin{minipage}[b]{\linewidth}\raggedright
Role
\end{minipage} \\
\midrule\noalign{}
\endhead
\bottomrule\noalign{}
\endlastfoot
\texttt{AvgCaseMls/Section3/} & 6,833 & CS87 resolution lower bounds \\
\texttt{AvgCaseMls/Foundation/} & 6,293 & Timed machine model, subprobabilities, reductions \\
\texttt{AvgCaseMls/Section4/} & 4,783 & Theorem 4.4, Example 4.1, Cook--Levin \\
Collapse + repair & \textasciitilde400 & Diagnostic results \\
Section 5 reductions & \textasciitilde3,000 & MLS-in, EMLS, FPILP \\
\end{longtable}

\textbf{Exposition snippets.} The \href{Exposition/}{\texttt{Exposition/}} directory holds runnable versions of every proof displayed in this document. Each file restates the theorem in context and either copies the proof (when it fits in ten lines) or applies it from the library. Check all snippets:

\begin{verbatim}
./scripts/check_exposition.sh
\end{verbatim}

\textbf{Palomar.} \texttt{Challenge.lean} remains the comparator surface for external verification; Palomar size caps are deferred until the canon is complete.

\begin{center}\rule{0.5\linewidth}{0.5pt}\end{center}

\hypertarget{references}{%
\section{References}\label{references}}

\begin{itemize}
\item
  \textbf{{[}Ajt96{]}} Ajtai, M. (1996). Generating hard instances of lattice problems. \emph{STOC}.
\item
  \textbf{{[}BDCGL89{]}} Ben-David, S., Chor, B., Goldreich, O., \& Luby, M. (1989). On the theory of average case complexity. \emph{STOC}.
\item
  \textbf{{[}CEM95{]}} Cox, J., Ericson, L., \& Mishra, B. (1995). The average case complexity of multilevel syllogistic. \emph{NYU Courant Institute Technical Report TR1995-711}.
\item
  \textbf{{[}COPE24{]}} Committee on Publication Ethics (COPE). (2024). Authorship and AI tools: COPE position statement. \url{https://publicationethics.org/guidance/cope-position/authorship-and-ai-tools}
\item
  \textbf{{[}Cur25{]}} Anysphere, Inc.~Cursor: AI-native code editor and agent environment. \url{https://cursor.com} (accessed 2025).
\item
  \textbf{{[}deM08{]}} de Moura, L., \& Bjørner, N. (2008). Z3: An efficient SMT solver. \emph{TACAS}.
\item
  \textbf{{[}DS77{]}} Davis, M., \& Schwartz, J. T. (1977). Metamathematical extensibility for theorem verifiers. \emph{NYU Technical Report}.
\item
  \textbf{{[}FOS80{]}} Ferro, A., Omodeo, E. G., \& Schwartz, J. T. (1980). Decision procedures for elementary sublanguages of set theory. \emph{CPAM}.
\item
  \textbf{{[}Gol79{]}} Goldberg, A. T. (1979). On the complexity of the satisfiability problem. \emph{NYU PhD Thesis}.
\item
  \textbf{{[}Gur91{]}} Gurevich, Y. (1991). Average case completeness. \emph{Journal of Computer and System Sciences}.
\item
  \textbf{{[}Lev86{]}} Levin, L. (1986). Average case complete problems. \emph{SIAM Journal on Computing}.
\item
  \textbf{{[}RS93{]}} Reischuk, R., \& Schindelhauer, C. (1993). Precise average case complexity. \emph{STOC}.
\item
  \textbf{{[}Ste23a{]}} Stevens, L. (2023). MLSS Decision Procedure. \emph{Archive of Formal Proofs}. \url{https://isa-afp.org/entries/MLSS_Decision_Proc.html}
\item
  \textbf{{[}Ste23b{]}} Stevens, L. (2023). Towards a Verified Tableau Prover for a Quantifier-Free Fragment of Set Theory. In Pientka, B., \& Tinelli, C. (eds.), \emph{Automated Deduction -- CADE 29}, LNAI 14132, 491--508. \url{https://doi.org/10.1007/978-3-031-38499-8_28}
\item
  \textbf{{[}SY92{]}} Schnorr, C. P., \& Yoshida, T. (1992). Average-case complexity of NP-complete problems. \emph{STOC}.
\item
  \textbf{{[}Sny90a{]}} Snyder, W. K. (1990). The SETL2 programming language. \emph{NYU Technical Report}.
\item
  \textbf{{[}ST01{]}} Spielman, D. A., \& Teng, S. H. (2001). Smoothed analysis of algorithms. \emph{STOC}.
\item
  \textbf{{[}VR92{]}} Venkatesan, R., \& Rajagopalan, S. (1992). Average case intractability of matrix and Diophantine problems. \emph{STOC}.
\end{itemize}

\end{document}